# Title: Direct observation of Dirac states in Bi$_2$Te$_3$ nanoplatelets by $^{125}$Te NMR

**Authors:** Wassilios Papawassiliou[1], Aleksander Jaworski[1], Andrew J. Pell[1*], Jae Hyuck Jang [2], Yeonho Kim[2], Sang-Chul Lee[2], Hae Jin Kim[2*], Yasser Alwahedi[3,7], Saeed Alhassan[3], Ahmed Subrati[3,4], Michael Fardis[5], Marina Karagianni[5], Nikolaos Panopoulos[5], Janez Dolinsek[6], and Georgios Papavassiliou[5*].

**Affiliations:**

[1]Department of Materials and Environmental Chemistry, Arrhenius Laboratory, Stockholm University, Svante Arrhenius vag 16 C, SE-106 91 Stockholm, Sweden,

[2] Electron Microscopy Research Center, Korea Basic Science Institute, 169-148 Gwahak-ro, Yuseong-gu, Daejeon 34133, Republic of Korea,

[3] Department of Chemical Engineering, Khalifa University, PO Box 2533, Abu Dhabi, United Arab Emirates,

[4]NanoBioMedical Centre, Adam Mickiewicz University, Wszechnicy Piastowskiej 3, 61-614 Poznań, Poland,

[5] Institute of Nanoscience and Nanotechnology, National Center for Scientific Research "Demokritos", 153 10 Aghia Paraskevi, Attiki, Greece,

[6]J. Stefan Institute and University of Ljubljana, Faculty of Mathematics and Physics, Jamova 39, SI-1000 Ljubljana, Slovenia,

[7]Center for Catalysis and Separation, Khalifa University of Science and Technology, P.O.Box 127788, Abu Dhabi, UAE

*andrew.pell@mmk.su.se, hansol@re.kbsi.kr, g.papavassiliou@inn.demokritos.gr.

**Abstract:** Detection of the metallic Dirac electronic states on the surface of Topological Insulators (TIs) is a tribune for a small number of experimental techniques the most prominent of which is Angle Resolved Photoemission Spectroscopy. However, there is no experimental method showing at atomic scale resolution how the Dirac electrons extend inside TI systems. This is a critical issue in the study of important surface quantum properties, especially topological quasiparticle excitations. Herein, by applying advanced DFT-assisted solid-state $^{125}$Te Nuclear Magnetic Resonance on Bi$_2$Te$_3$ nanoplatelets, we succeeded in uncovering the hitherto invisible NMR signals with magnetic shielding influenced by the Dirac electrons, and subsequently showed how Dirac electrons spread and interact with the bulk interior of the nanoplatelets.

**One Sentence Summary:** Advanced 2D $^{125}$Te NMR techniques on Bi$_2$Te$_3$ nanoplatelets uncover surface signals with NMR shielding due to the Dirac electron orbital motion. (125 characters)

**Main Text:**

In the presence of spin-orbit coupling (SOC), the spin $\hat{s}$ and the orbital angular momentum $\hat{l}$ of the electron lose their time invariance; it is the total angular momentum of the electron $\hat{j} = \hat{s} + \hat{l}$ that preserves it. This fundamental property of electrons is the playground of a number of fascinating phenomena, such as the Dirac edge states in topological insulators (TIs) (*1*), the quantum spin Hall effect (*2*), and the formation of Majorana fermions (*3-5*). In case of three-dimensional TIs, SOC forces the surface electrons to form helical spin structures, wrapping an odd number of massless Dirac cones, with the simplest and most-studied systems being the Bi$_2$Se$_3$ and $Bi_2Te_3$ tetradymides. Angle resolved photoemission spectroscopy (ARPES) experiments in combination with theoretical studies have shown that these systems acquire a single Dirac Cone and large band gap (6, 7), thus providing an ideal platform for studying topological quantum properties. However, despite the simplicity of their topological edge states, important collective effects of Dirac electrons, such as the Majorana zero mode at the interface of TI/SC heterostructures (3-5), the excitonic superfluid condensate (8), or the propagation of chiral spin waves on topological surfaces (9) remain to a great extent experimentally unexplored.

Since many topological properties depend on the way that the spin of the Dirac electrons couples with their orbital motion and how this interaction propagates through the crystal, an experimental probe sensitive to both the spin and orbital motion of the Dirac electrons is crucial in the efforts to further understand the physics of topological materials. Nuclear magnetic resonance (NMR) appears to fulfill these requirements as the nuclear magnetic shielding, and consequently the NMR frequency shift, depends on the spin and orbital magnetic susceptibility at the position of each resonating nucleus. Specifically, the total NMR Knight shift can be expressed as $K=K_{FC}+K_D+K_{ORB}$ , where the first two terms are the Fermi-contact and spin-dipolar terms originating from the electron spin polarization at the Fermi level, and the third is the orbital term generated by the orbital currents of the Dirac electrons (10, 11). According to recent theoretical calculations on TI's the orbital term $K_{ORB}$ from the Dirac electrons dominates over $K_{FC}$ and $K_D$, and induces large negative shifts, and very short spin-lattice relaxation times $T_1$ (11). However, until now experimental $^{125}Te$ NMR studies on $Bi_2Te_3$ nanoparticles (12) and microcrystalline (bulk) powders (13) have given contradicting results.

In this report, implementing novel DFT-assisted broadband solid-state NMR methods with $C_s$-corrected scanning transmission electron microscopy (STEM) on $Bi_2Te_3$ nanoplatelets of high crystallinity, we demonstrate the first direct NMR detection of Dirac topological states. Figures 1A and S1-S3 illustrate the excellent quality of the nanoplatelets, which exhibit perfect hexagonal shapes with sharp edges, and average diameter of 600 nm, and a mean thickness of 10 nm. The structural characteristics were examined at atomic scale by means of High Angle Annular Dark/Bright Field HAADF/ABF imaging, as presented in Figures 2A and S3. Since the intensity of the HAADF images is proportional to the atomic number Z2, atomic columns in the HAADF/ABF image were labeled according to the Bi, Te(1) and Te(2) atomic sites by red, blue, and magenta dots respectively (14). The Bi and Te atoms are organized in well-defined quintuple atomic layers comprising of five covalently bonded atomic sheets of alternating Bi and Te atoms, i.e. Te(1)-Bi-Te(2)-Bi-Te(1), that are bound to each other by van der Waals interactions, as seen in the intensity profiles in Figures 2B and S3.

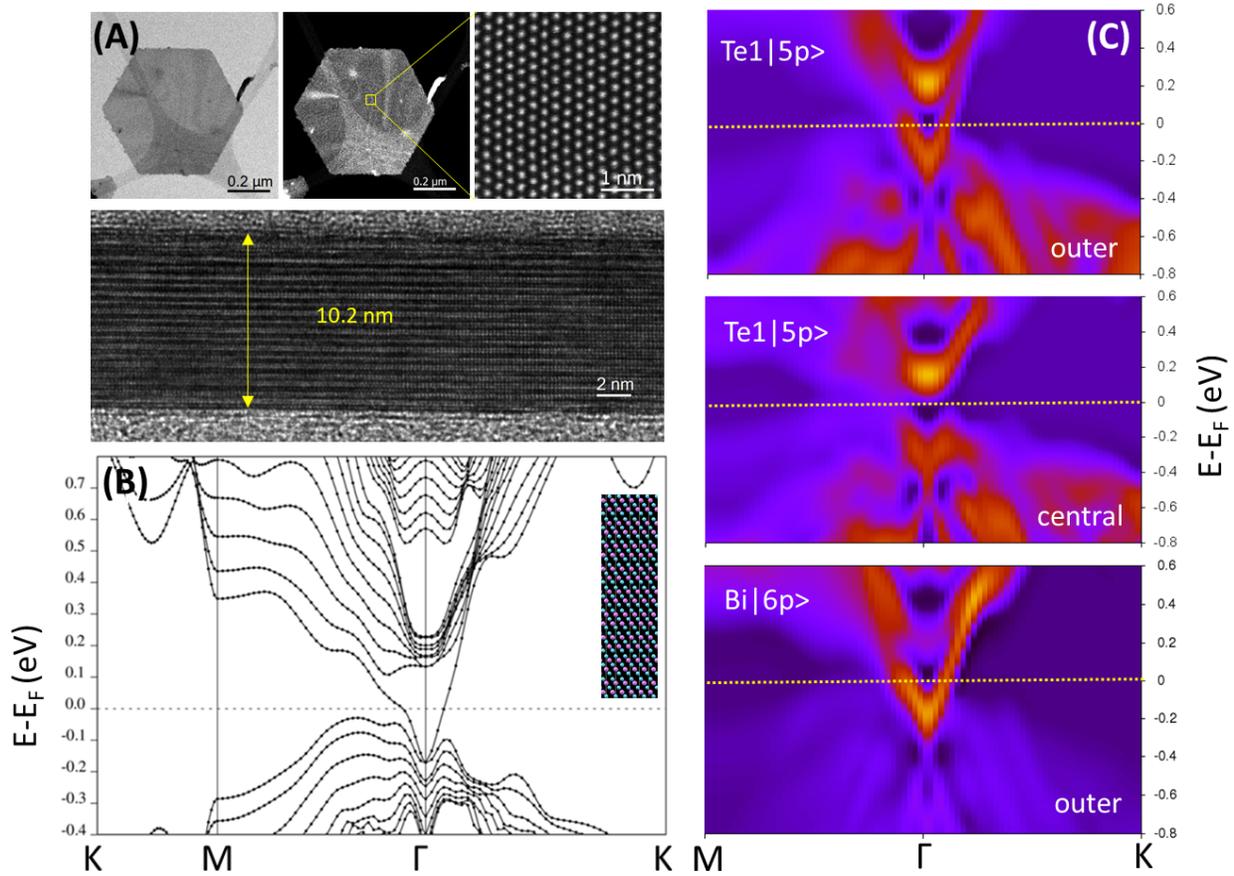

**Fig. 1. Band structure analysis and Dirac states of stoichiometric Bi₂Te₃ nanoplatelets. A.** High resolution TEM and HAADF image of Bi₂Te₃ nanoplatelets (top view and cross-section). **B.** Band structure of a 9-quintuplets-thick (∼10 nm) Bi₂Te₃ slab. **C.** The projected *k*-resolved DOS of the Te(1) |5*p*⟩ and Bi |6*p*⟩ orbital states at the outer (edge) quintuplets, and the Te(1) |5*p*⟩ orbital states at the central quintuplet. Dirac states are observed only at the edge quintuplets (more details in Figure S6).

In order to examine the way in which the Dirac states propagate through the nanoplatelets and how this feature is encoded into the NMR Knight shift, DFT calculations were carried out on a Bi₂Te₃ slab comprising 9 quintuplets, corresponding to the mean thickness of the Bi₂Te₃ nanoplatelets. Details on the DFT calculations are given in the SI. Figures 1C, and S4-S6 show that Dirac states are defined mainly by the Te(1) |5*p*⟩ and Bi |6*p*⟩ states of the terminating quintuplets. Furthermore, the density of the Dirac states is substantially reduced in the central region of the nanoplatelets, as shown in the middle panel of Figure 1C and Figure S6. It is therefore expected that the NMR signals from the surface of the nanoplatelets will be shifted with respect to the signals from the bulk (interior), because the Dirac electrons are predicted to induce large negative orbital Knight shifts *(11)*. Despite this expectation, until now efforts to detect the Dirac states through ¹²⁵Te NMR have not been successful, either with static or with magic-angle-spinning (MAS) one-dimensional (1D) NMR methods. The main reason is the large Knight shift anisotropy, which gives rise to broad unresolved NMR signals that are also difficult to excite.

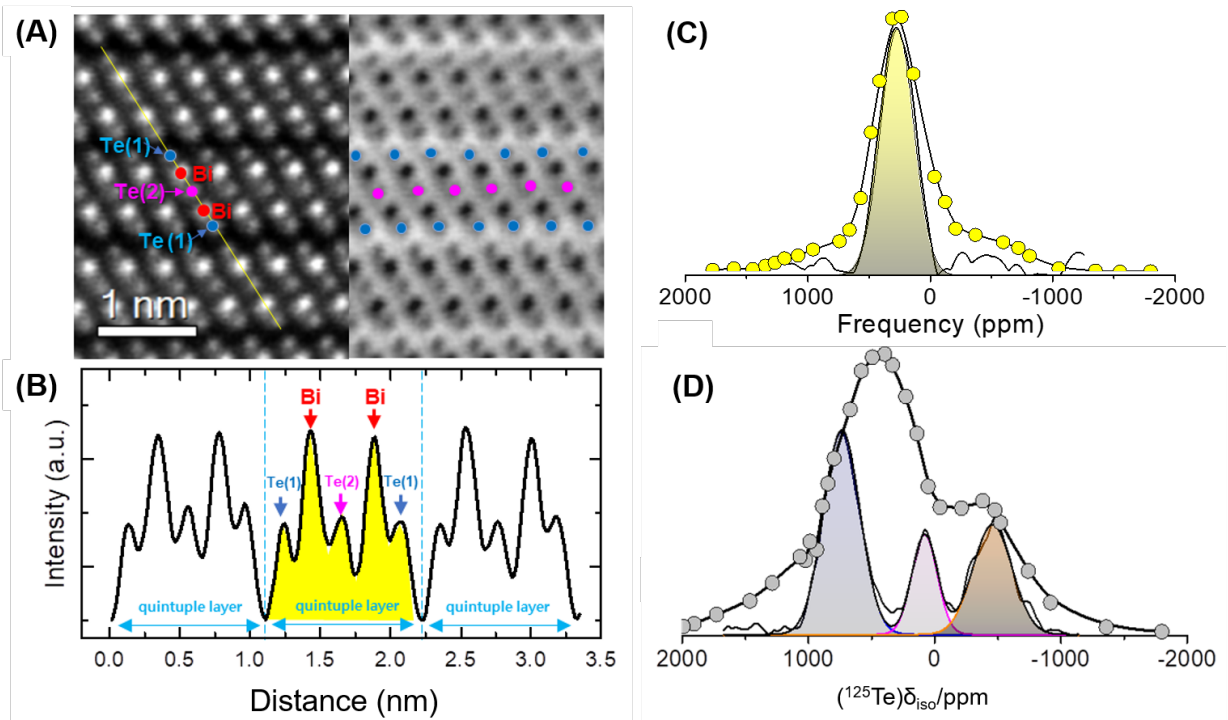

**Fig. 2. Atomic scale TEM analysis and 1D $^{125}$Te NMR. A.** Cross-sectional HAADF-ABF images of a Bi$_2$Te$_3$ nanoplatelet. Red, blue and magenta dots indicate the Bi, Te(1) and Te(2) columns in the quintuplets. **B.** The intensity profile from the cross section (yellow line) in the HAADF image clarifies the atomic positions of the Bi, Te(1), and Te(2) atoms. **C.** $^{125}$Te static frequency sweep NMR (yellow circles) and the isotropic projection of the 2D $^{125}$Te aMAT NMR at 14 kHz MAS (solid line) of bulk Bi$_2$Te$_3$. **D.** $^{125}$Te static frequency sweep NMR (grey circles) and the isotropic projection of the 2D $^{125}$Te aMAT at 30 kHz MAS of a Bi$_2$Te$_3$ nanoplatelets sample.

The shortcomings of 1D NMR for acquiring well-resolved spectra is clearly seen in Figures 2C and 2D, which compare the static frequency-sweep $^{125}$Te NMR spectra of the Bi$_2$Te$_3$ microcrystalline (bulk) and nanoplatelet samples. Both spectra exhibit different overall profiles, but with similar features; each is characterized by a central peak at a shift of 250 ppm (microcrystals) and 480 ppm (nanoplatelets), respectively, with a tail at higher shift, and a shoulder at negative shift. In case of the microcrystalline material these features have been explained as originating from the presence of two overlapping signals shifted relative to each other *(13)*: one strong narrow resonance from Te(1), and a broader asymmetric resonance at a shift of -400 ppm from Te(2). Implementation of advanced 1D solid-state MAS NMR methods, such as the double adiabatic echo (DAE) experiment did not improve the resolution, as seen in Figure S8C of the SI.

In order to resolve these individual $^{125}$Te NMR signals, and to identify those from the topological edge states, the 2D adiabatic magic-angle turning (aMAT) NMR experiment was implemented (Figure 3A), which in the indirect dimension provides the isotropic NMR shifts free from spectral broadening due to any kind of anisotropy *(15, 16)*. Details on the implemented NMR techniques are provided in the SI. The effectiveness of the method is in evidence in Figures 2C and 2D, which overlay the isotropic $^{125}$Te NMR projections of the 2D aMAT of both the microcrystalline and nanoplatelet samples onto the 1D frequency-sweep spectra. In the case of the microcrystalline sample a single broad resonance is observed at 250 ppm, which comprises the two overlapping Te(1) and Te(2) NMR signals. No isotropic signal component is observed at negative shifts, as is also clearly shown in Figure S7. In case of the nanoplatelets the isotropic "bulk" NMR signal resolves into two distinct components at shifts of 765 and 93 ppm with a ratio of integrals of 2:1, corresponding to Te(1) and Te(2) respectively, whereas a third distinct signal component is observed at a markedly different shift of -452 ppm, which is assigned to the topological edge states. This signal appears reproducibly in different nanoplatelet samples, including those exhibiting some oxidation of the surface, as observed in Figure S8, which is an important finding for many applications. Furthermore, the shift difference in the isotropic bulk NMR signals between the two systems reflects differences in the electron/hole doping in the different particle morphologies *(17)*, and in general in distribution of the conduction electrons across the nanoplatelets.

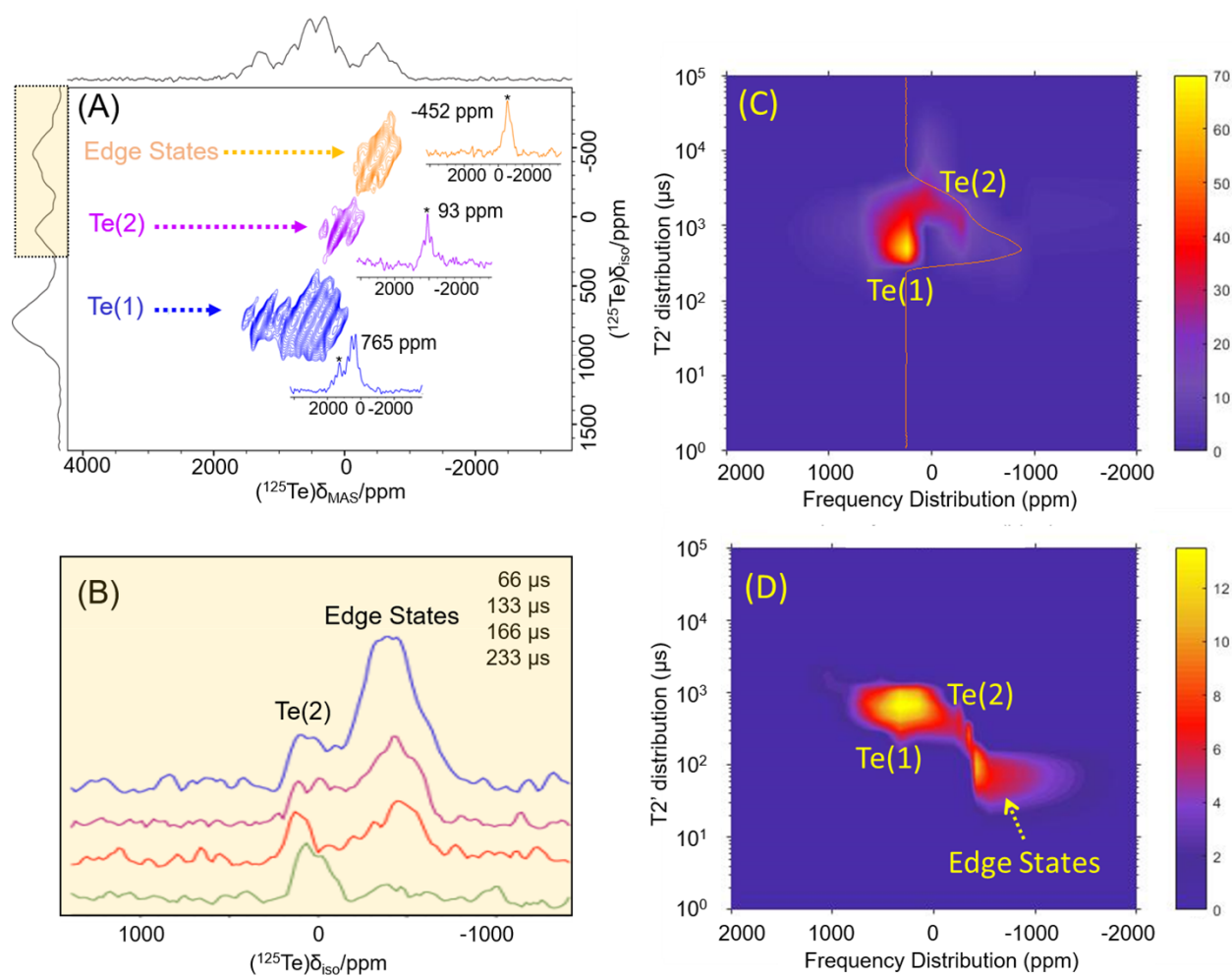

**Fig. 3. $^{125}$Te aMAT NMR and $T_2'$ dephasing analysis of the Dirac edge states. A.** 2D $^{125}$Te aMAT NMR spectrum of $Bi_2Te_3$ nanoplatelets. Blue and magenta colour contours indicate signals from the bulk-like interior of the nanoplatelets, while the orange colour contours show signals from the surface Te sites, shielded by the motion of the Dirac electrons. **B.** The expanded isotropic

projections of $^{125}$Te MAT NMR spectrum acquired at four different aMAT constant time showing differential $T_2'$ dephasing of the different Te environments. **C.** The $^{125}$Te NMR $T_2'$ distribution as a function of the resonance frequency of microcrystalline (bulk) $Bi_2Te_3$. The orange colour cross section shows the $T_2'$ distribution at frequency 250 ppm. **D.** The $^{125}$Te NMR $T_2'$ distribution with respect to the resonance frequency of the $Bi_2Te_3$ nanoplatelets.

We note that, because of the long adiabatic pulses (33.33 μs) *(16)*, aMAT signals with very short spin-spin relaxation times $T_2$ may be partially wiped out. The way that nuclear spin coherences dephase across the spectrum is shown in Figure 3C and 3D, which display the $T_2'$ distribution vs. shift for both the microcrystalline and nanoplatelet samples, acquired by 1D inverse-Laplace-transform Carr-Purcell-Meiboom-Gill (CPMG) spin-echo pulse trains. In addition to the inherent $T_2$ relaxation, $T_2'$ dephasing includes coherent signal decay due to the extended nuclear dipolar coupling network across the particles *(19)*. Experimental details on the CPMG inversion are presented in the SI. In the case of the microcrystalline sample the two signals corresponding to Te(1) and Te(2) are resolved, with Te(2) showing a highly anisotropic frequency distribution. In case of the nanoplatelets, the third strong signal component at -500 ppm, exhibits significantly shorter $T_2'$ times, marking the enhanced relaxation induced by the Dirac electrons *(11)*. To test whether such short $T_2'$ is influencing the signal intensity in the aMAT spectrum, the isotropic shift projections of standard MAT experiments were obtained at different shift evolution times is shown in Figure 3B *(15, 16, 18)*. The large signal intensity at -500 ppm at short evolution times indicates that a large proportion of the atomic layers in the nanoplatelets have a surface-like electronic structure, and its rapid dephasing at longer evolution times relative to the other signals due to the nanoplatelet interior confirms the difference in dephasing times shown in Figure 1D.

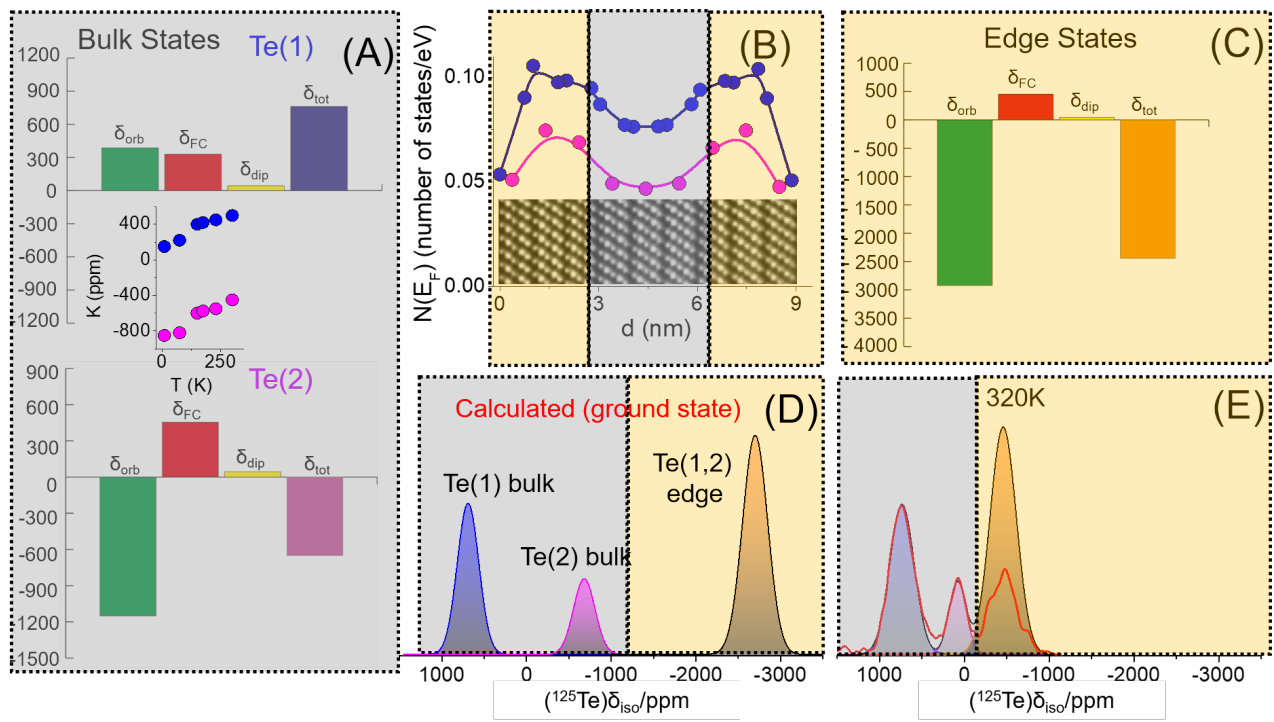

**Fig. 4. DFT analysis of the $^{125}$Te NMR Knight shifts in $Bi_2Te_3$ nanoplatelets.** Grey/yellow shaded areas refer to bulk/topological states. **A.** Calculated $^{125}$Te orbital, Fermi contact (spin), and dipolar terms of the Knight shift in the presence of SOC of microcrystalline $Bi_2Te_3$. The inset

shows the temperature dependence of the Knight shift according to ref. *(13)*. **B.** The Te(1) and Te(2) DOS at the Fermi level $N(E_F)$, across a 9-quintuplet (~9 nm) $Bi_2Te_3$ slab. **C.** Calculated $^{125}Te$ orbital, Fermi contact (spin), and dipolar terms of the Knight shift in the presence of SOC of a 5-quintuplet $Bi_2Te_3$ slab. **D.** Simulation of the $^{125}Te$ NMR signal of a 9-quintuplet slab, determined by combining the calculated Knight shifts. The peak intensity ratios were calculated according to the number of Te sites in the grey/yellow shaded areas in panel (B). Theoretical peaks were convoluted with a Gaussian function to simulate the linewidths of the experimental spectra. **E.** Experimental isotropic $^{125}Te$ NMR spectrum of the nanoplatelet spectrum. The NMR signal from the edge states in the yellow shaded area was corrected to account for differential $T_2'$ signal dephasing.

To confirm the assignment of the $^{125}Te$ aMAT spectra, DFT calculations of the NMR shifts were carried out, using the full-potential linearized augmented plane-wave method, as implemented in the Wien2k DFT software package. Figures 4(A,C) show the calculated orbital, Fermi-contact (spin), and spin-dipolar terms of the Knight shift in the presence of SOC, of both microcrystalline $Bi_2Te_3$ and a 5-quintuplets-thick slab representing the nanoplatelets (with thickness $d$~5 nm). The latter was selected because it allows tractable DFT calculations with satisfactory $k$-grid, while Dirac electron states remain gapless *(20)*. The orbital terms were referenced according to $\delta_{iso\text{-}ORB} = \sigma_{ref}$ -$\sigma_{iso\text{-}ORB}$ *(21)*, with the reference shielding set to $\sigma_{ref}$ =2370 ppm, according to Figure S9 in the SI. In case of the microcrystalline material the total isotropic Knight shifts were found to be +720 ppm for Te(1) and -660 ppm for Te(2), as shown in Figures 4(A,D). However, in the case of the 5-quintuplet-thick slab the isotropic orbital Knight shifts of Te(1) and Te(2) coalesce to a single resonance, acquiring strong negative shift, $\delta_{iso\text{-}ORB}$~-2670 ppm. This is explainable by the fact that in very thin nanoplatelets ($d \leq 5$ nm), Dirac electrons spread almost uniformly across the slab (Figure S5), inducing a strong negative Knight shift via their orbital currents, in agreement with theoretical predictions *(11)*. Notably, by increasing the slab thickness the system splits into an interior region that occupies the center of the slab, and which is sandwiched between two "topological" surface regions. This is shown in Figure 4B, which displays the Te(1) and Te(2) DOS at the Fermi level across a 9-quintuplets slab. At the center of the nanoplatelet the DOS drops significantly and a band gap begins to open, as shown in Figures 2C and S6. On the basis of these results, the calculated ground state $^{125}Te$ NMR spectrum of a 9-quintuplet-thick slab was depicted (Figure 4D), and compared with the experimental isotropic $^{125}Te$ NMR spectrum of the nanoplatelets (Figure 4E). In the latter case, the shown signal intensities have been corrected by considering the $T_2'$ dependence of the signal on the interpulse time intervals of the aMAT experiment.

Bearing in mind the strong negative NMR frequency shift by decreasing temperature *(11, 13)*, a nice correlation between the theoretical and experimental spectra is witnessed. This is strong evidence that the isotropic $^{125}Te$ NMR signal component at -452 ppm belongs to the surface NMR signal shielded by the Dirac electrons. In this perspective, the intensity ratio of the surface-to-bulk NMR signals provides the mean volume of the nanoplatelets that is occupied by the Dirac electrons, whereas $T_1$ (Figure S11) and $T_2$' relaxation measurements on the pertinent peaks highlight the interaction of the Dirac electrons with the bulk interior. It is anticipated that these results open up new ways to study of quantum topological phenomena.

**Acknowledgments:** W.P., A.J., and A.J.P. were supported by the Swedish Research Council (project no. 2016-03441). M.K. and G.P. acknowledge support by the project MIS 5002567, implemented under the "Action for the Strategic Development on the Research and Technological Sector", funded by the NSRF 2014-2020 and co-financed by the European Union and Greece. Part of the DFT work was performed using computational resources of the Research Computing Department at Khalifa University.

# Supplementary Materials for

Direct observation of Dirac states in $Bi_2Te_3$ nanoplatelets by [125]Te NMR

Wassilios Papawassiliou[1], Aleksander Jaworski[1], Andrew J. Pell[1*], Jae Hyuck Jang [2], Yeonho Kim[2], Sang-Chul Lee[2], Hae Jin Kim[2*],Yasser Alwahedi[3,7], Saeed Alhassan[3], Ahmed Subrati[3,4], Michael Fardis[5], Marina Karagianni[5], Nikolaos Panopoulos[5], Janez Dolinsek[6], and Georgios Papavassiliou[5*].

**Affiliations:**
[1]Department of Materials and Environmental Chemistry, Arrhenius Laboratory, Stockholm University, Svante Arrhenius vag 16 C, SE-106 91 Stockholm, Sweden,

[2] Electron Microscopy Research Center, Korea Basic Science Institute, 169-148 Gwahak-ro, Yuseong-gu, Daejeon 34133, Republic of Korea,

[3] Department of Chemical Engineering, Khalifa University, PO Box 2533, Abu Dhabi, United Arab Emirates,

[4]NanoBioMedical Centre, Adam Mickiewicz University, Wszechnicy Piastowskiej 3, 61-614 Poznań, Poland

[5] Institute of Nanoscience and Nanotechnology, National Center for Scientific Research "Demokritos", 153 10 Aghia Paraskevi, Attiki, Greece,

[6]J. Stefan Institute and University of Ljubljana, Faculty of Mathematics and Physics, Jamova 39, SI-1000 Ljubljana, Slovenia,

[7]Center for Catalysis and Separation, Khalifa University of Science and Technology, P.O.Box 127788, Abu Dhabi, UAE

*Correspondence to: andrew.pell@mmk.su.se, hansol@kbsi.re.kr, g.papavassiliou@inn.demokritos.gr

**This PDF file includes:**

Materials and Methods
Supplementary Text
Figures S1 to S11

**Materials and Methods**

<u>Materials</u>

Bi$_2$Te$_3$ nanoplatelets were synthesized following a solvothermal approach. Specifically, 1 mmol BiCl$_3$ and 1.5 mmol Na$_2$TeO$_3$ were dispersed in 15 mmol of an alkaline solution (NaOH), and 1.16 M polyvinylpyrrolidone (PVP, M$_w$ = 40,000 g/mol Da) were dissolved in 40 mL of ethylene glycol. The mixture was magnetically stirred until it turned highly translucent, then it was transferred and sealed into a Teflon-lined stainless-steel autoclave (capacity of 80 mL). The sealed autoclave was put into an oven at 180 °C for 36 hrs and cooled to room temperature. The resulting products were collected by repeated centrifugations, and subsequently washed with distilled water and ethanol, two times each, and finally vacuum dried overnight at 90 °C for further characterization.

<u>Methods</u>

<u>*NMR*</u>

The $^{125}$Te MAS experiments of the Bi$_2$Te$_3$ microcrystalline material were performed with a 4 mm HXY triple-resonance probe, at 14 kHz MAS on a Bruker 400 Avance-III spectrometer operating at a $^{125}$Te Larmor frequency of 126.23 MHz. Spectral acquisition was done with a double adiabatic spin-echo sequence with a 2.5 μs 90° excitation pulse length, corresponding to an RF field of 100 kHz, followed by a pair of rotor-synchronized short high-power adiabatic pulses (SHAPs) *(1)* of 71.43 μs length and a 5 MHz frequency sweep. 32768 transients were acquired with a recycle delay of 2 seconds. For the separation of the isotropic shift and chemical shift anisotropy, which is of significant magnitude in heavy spin-1/2 nuclei *(2)* and strong electron correlated systems, the adiabatic magic-angle-turning (aMAT) *(3)* pulse sequence was employed, which consists of a π/2 excitation pulse followed by 6 refocusing SHAP π-pulses (Figure S7). The same SHAPs as in the double adiabatic spin-echo sequence were used. The aMAT evolution time was 71.43 μs, excluding the length of the SHAPs equal to one rotor period. In the indirect dimension 32 increments were collected, in which the spectral width was 224.21 kHz. 2560 transients were acquired for each increment. Each transient had a recycle delay of 2s. The $^{125}$Te MAS spectra of the Bi$_2$Te$_3$ nanoplatelets were acquired on a Bruker 400 Avance-III spectrometer operating at Larmor frequency of 126.23 MHz with 2.5 mm HX probe, at 30 kHz MAS. For the acquisition of the double adiabatic echo and the aMAT spectra, rotor synchronized SHAPs sweeping through 5 MHz in 33.33 μs were employed with RF field amplitude of 160 kHz. For the double-adiabatic-echo experiment, 40960 scans with a recycle delay of 1s were sufficient for the acquisition of the spectrum. For the aMAT spectra, the same SHAPs were used with an evolution time of 66.66 μs, excluding the length of the SHAPs, which is equivalent to two rotor periods. Two aMAT spectra were acquired, one with a recycle delay of 70 ms and implementation of 48 increments in the indirect dimension, with a spectral width 359.72 kHz and 6144 scans per increment; The second with the same parameters, except a recycle delay of 4 s and 1280 scans per increment. For the MAT sequence, five π-pulses of l.5 μs were employed with a recoupling time of 66.66 μs, 133 μs, 166 μs and 233 μs respectively, excluding the π-pulses. All chemical shifts were referenced to TeO$_2$ *(4)*.

The frequency-sweep $^{125}$Te NMR spectra were acquired on a home-built NMR spectrometer under static conditions, operating at Larmor frequency of 126.23 MHz. For the spin-lattice relaxation time $T_1$ and the coherence lifetime $T_2'$ experiments a π/2-t-π/2 saturation recovery pulse sequence and a Carr-Purcell-Meiboom-Gill (CMPG) pulse sequence π/2-τ-{π-2τ-π-...-π-2τ-π}$_{300}$ with a train of 300 π pulses were implemented, respectively. The $T_1$ and $T_2'$ distribution analysis was performed by applying a non-negative Tikhonov Regularization Algorithm *(5, 6)*, as described below.

*Electron Microscopy*

Scanning electron microscopy (SEM) images were recorded on a Hitachi S4800 microscope while the size distribution of the $Bi_2Te_3$ nanoplatelets were obtained by measuring 100 randomly selected $Bi_2Te_3$ platelets in the SEM image. Scanning transmission electron microscopy (STEM) images and their energy-dispersive X-ray (EDX) elemental mappings were measured using a JEOL JEM-2100F.

High resolution TEM and STEM images were acquired using a Jeol ARM200 with probe $C_s$-corrector, operated at 200 kV. Cross sectional TEM samples were prepared by FIB in KBSI (Quanta 3D FEG, FEI). Atomic-coordinates analysis from HAADF images was performed via intensity refinement method.

The atomic coordinates analysis was performed with the following steps: (i) HAADF image normalization, (ii) Laplacian of Gaussian filtering, (iii) image erosion, (iv) atom position detection using circular pattern matching, (v) pattern matching errors correction, (vi) center position of atoms correction, (vii) computation of the average pixel intensity around the center of atoms, and finally (viii) two kind of markers based on intensity values has been depicted on the HAADF image, so that the coordinates of Bi and Te are clearly identified from each other.

*X-ray Diffraction*

The XRD spectra were recorded with an analytical PANalytical X'Pert PRO powder diffractometer. The sample was mounted on a zero-background holder and scanned by using Cu-Kα radiation ($\lambda = 1.5418$ Å) with the following experimental conditions: applied voltage of 40 kV, intensity of 30 mA, angular range (2θ) $5 - 80°$ and 0.03 steps/s. Rietveld refinement of obtained powder XRD pattern was carried out using the FULLPROF program software. Refined parameters include: overall scale factor, background (BGP), lattice parameters, atomic positions and orientation.

*Density Functional Theory (DFT) calculations*

DFT calculations were carried out with the QUANTUM ESPRESSO package *(7)* on a $Bi_2Te_3$ slab comprised of 9 quintuplets, acquiring the mean thickness of the $Bi_2Te_3$ nanoplatelets. The slab surface was set according to the top-view HAADF image in Figure 1A, along the (001) plane. Calculations were performed on the basis of the Perdew_Burke_Ernzerhof (PBE) type generalized gradient approximation. For the Brillouin zone integrations we used a 11x11x1 Monkhorst-Pack **k**-point mesh, and the kinetic energy cutoff was fixed to 800 eV. The lattice constants were acquired by the Rietveld refinement of the XRDs ($a=b=$ 4.395 Å, $c=$ 29.830 Å). Spin-orbit effects were treated self-consistently using fully relativistic Projector Augmented Wave (PAW) pseudopotentials *(8)*.

NMR Knight Shift calculations were performed by using the full-potential linearized augmented plane-wave method, as implemented in the Wien2k DFT software package *(9)*. The spin-orbit interaction was considered in a second variational method. Calculations were performed with and without spin-orbit coupling on two different atomic configurations; bulk $Bi_2Te_3$, and a 5 quintuplets $Bi_2Te_3$ slab with thickness ~5 nm. The k-mesh convergence was checked up to 100,000 points for the bulk materials and up to 5,000 points for the slab.

**Supplementary Text**

*Morphological and compositional features of pristine $Bi_2Te_3$ nanoplatelets (Figures S1-S3)*

Figure S1 presents the morphological characteristics of pristine $Bi_2Te_3$ nanoplatelets. Hexagonal (0001) facets can be seen in Figure S1(A). The Rietveld refinement presented in Figure S1(B) clearly shows a plausible match with the employed theoretical model with no interference from the selected background (BGP). The diffraction pattern can be indexed to JCPDS: 15-0863. The inset shows the high quality and credibility of the refinement. The stoichiometry maps in Figure S1(C) show homogeneous atomic distribution of constituents with stoichiometric higher Te intensity. The EDX results show Bi and Te signals with very good match to the ideal stoichiometry, i.e. 40.7 at.% Bi and 59.3 at.% Te. The size distribution is quite narrow with the majority of nanoplatelets sized around 600 nm (Figure S1(D)).

Figure S2(A) shows a low magnification High-Angle Annular Dark-Field (HAADF) image and a cross-sectional TEM image of two nanoplatelets (Figure S2(B,C)). The average size of the fabricated nanoplatelets is ~10 nm. The High resolution cross-sectional HAADF image of $Bi_2Te_3$ along the [210] direction in Figure S3 shows the excellent stoichiometry and crystal structure of the synthesized samples. It is furthermore noticed that the excellent Rietveld analysis of the XRD pattern and the X-ray energy dispersion spectroscopy results presented in Figure S1 confirm the absence of any Te-based impurity phase.

*Density of States (DOS) and k-resolved band structure analysis of $Bi_2Te_3$ nanoplatelets (Figures S4-S6)*

The DOS of a 5-quintuplets $Bi_2Te_3$ slab (25 atoms in the supercell) and of a 9-quintuplet $Bi_2Te_3$ slab (45 atoms in the supercell) were calculated in order to examine the role of the various atomic orbitals to the Dirac electron states and subsequently to the [125]Te NMR Knight shifts. Each quintuplet comprises of a Te(1)-Bi-Te(2)-Bi-Te(1) atomic arrangement, whereas successive quintuplets are bonded to each other with van der Waals forces.

Figure S4 shows the total DOS of both $Bi_2Te_3$ slabs with and without Spin Orbit Coupling (SOC). In the absence of SOC, a 200 meV energy gap opens between the Valence Band Maximum (VBM) and the Conduction Band Minimum (CBM). When SOC is turned on, the gap is closing and a finite DOS is crossing the Fermi level, which looks to be almost the same in both atomic arrangements. In the case of slabs thinner than 5 quintuplets an energy gap was observed to open at the Dirac point.

Figure S5 shows Te(1) and Te(2) DOS at the Fermi level $N(E_F)$, across a 5-quintuplet (~5 nm) $Bi_2Te_3$ slab. For nanoplatelets with thickness d≤5 nm Dirac electrons extend uniformly across the nanoplatelets

Figure S6 presents the calculated k-resolved projected DOS of the outermost and central quintuplets of the 45 atoms $Bi_2Te_3$ slab. It is clearly seen that atoms in the outermost quintuplets have sufficiently higher contribution to the Dirac electron DOS. Evidently, by increasing the slab thickness the projected DOS of atoms in the central quintuplets attains a "bulk-like" character.

*[125]Te aMAT NMR of bulk $Bi_2Te_3$ (Figure S7)*

To determine the isotropic Knight shift of the two non-equivalent Te sites we employed the aMAT pulse sequence, which is a constant period pulse train of six short high-power adiabatic pulses (SHAPs) following an initial 90° excitation pulse, as shown in the inset of Figure S7. The main figure highlights the strength of the aMAT experiment, being able to eliminate anisotropies unveiling the presence of just a single broad Knight shift, containing the two overlapping, non-equivalent Te sites.

*[125]Te aMAT NMR of slightly oxidized $Bi_2Te_3$ nanoplatelets (Figure S8)*

Figure S8 showcases the excellent resolution that can be achieved when performing the aMAT experiment on the ultrathin $Bi_2Te_3$ nanoplatelets. Four well-resolved environments are observed, of which three are attributed to the inequivalent tellurium sites and the edge states, respectively of $Bi_2Te_3$, and the fourth is attributed to a small number of surface oxidized nanoplatelets (1410 ppm). It is noticed that slight oxidization is not influencing Dirac edge states.

<u>*Correlation between the experimental isotropic chemical shift and the calculated isotropic magnetic shielding of prototype Te-containing compounds (Figure S9)*</u>

In order to obtain the reference isotropic $^{125}$Te magnetic shielding $\sigma_{ref}$, NMR calculations were performed on three prototype materials in the presence of spin-orbit coupling. The data in Figure S9 are plots of the calculated isotropic magnetic shielding as function of the experimental isotropic chemical shifts; the latter were taken from refs. *(10, 11)*. Line is fit to the plot according to formula $\sigma_{calc} = 2370.1 - 0.907\delta_{iso}$. On the basis of this result the calculated reference shielding used in this article was set equal to $\sigma_{ref} = 2370.1$.

<u>*$^{125}$Te NMR $T_2'$ distribution (Figure S10)*</u>

The upper panel in Figure S10 shows schematically the typical CPMG pulse sequence. Experimental spin-echo decay curves were acquired by recording the intensity of consecutive spin echoes. CPMG spin echoes decay under the effect of the inherent $T_2$ relaxation, the presence of coherent dephasing, and eventually electron-nuclear interactions coupled with the CPMG pulse train. The overall effective dephasing time constant is defined as $T_2'$. In case of the $Bi_2Te_3$ nanoplatelets the presence of spin-diffusion was confirmed by recording the spin echo decay curves at different delay times $2t_D$.

The acquired CPMG spin echo decay trains were made of 300 echoes with interecho distance $2t_E = 40$ μs (exemplary experimental spin echo decay is shown as black line in the main panel). The inset shows the $g(T_2')$ distribution after inverting the spin-echo decay curve. The blue line in the main panel is the theoretical fit on the experimental data.

In order to acquire the $^{125}$Te NMR spin-spin relaxation time $T_2'$ distribution function $g(T_2')$, the experimental CPMG spin-echo decay curves were modelled with a Fredholm integral equation of the first kind *(5,6)*, $\frac{M(t)}{M(0)} = \int_0^{+\infty} k_0(T_2', t) g(T_2') d(log_{10}T_2')$, where $\frac{M(t)}{M(0)}$ is the normalized CPMG spin echo decay and $k_0(T_2', t) = exp\left(-\frac{t}{T_2'}\right)$. This equation can be transformed in a vector matrix notation to $\boldsymbol{M} = \boldsymbol{K_0}\boldsymbol{g}$, which in turn can be inverted to obtain the $g(T_2')$ distribution function *(5,6)*. In the present work, the inversion was achieved by implementing a modified non-negative Tikhonov regularization algorithm *(6)*.

The contour plot in Figures 3C and 3D of the main article are made out of 30 $g(T_2')$ curves, acquired at 30 consecutive resonance frequencies, covering the whole $^{125}$Te NMR spectra. Soft pulses were used so that each time a narrow frequency bandwidth was irradiated.

<u>*$^{125}$Te NMR $T_1$ distribution vs. Frequency (Figure S11)*</u>

The spin-lattice relaxation time distribution function $g(T_1)$ was obtained in a similar way as $g(T_2')$, by replacing the kernel $k_0(T_2', t)$ in the integral equation with $k_0(T_1, t) = 1 - exp\left(-\frac{t}{T_1}\right)$. Figure S11 shows the $^{125}$Te NMR $g(T_1)$ of the nanoplatelet and bulk systems. In case of the nanoplatelets, the NMR frequencies are very close to the three isotropic resonances of the aMAT spectra. A striking difference in the $T_1$ values between the bulk and the nanoplatelets systems is observed.

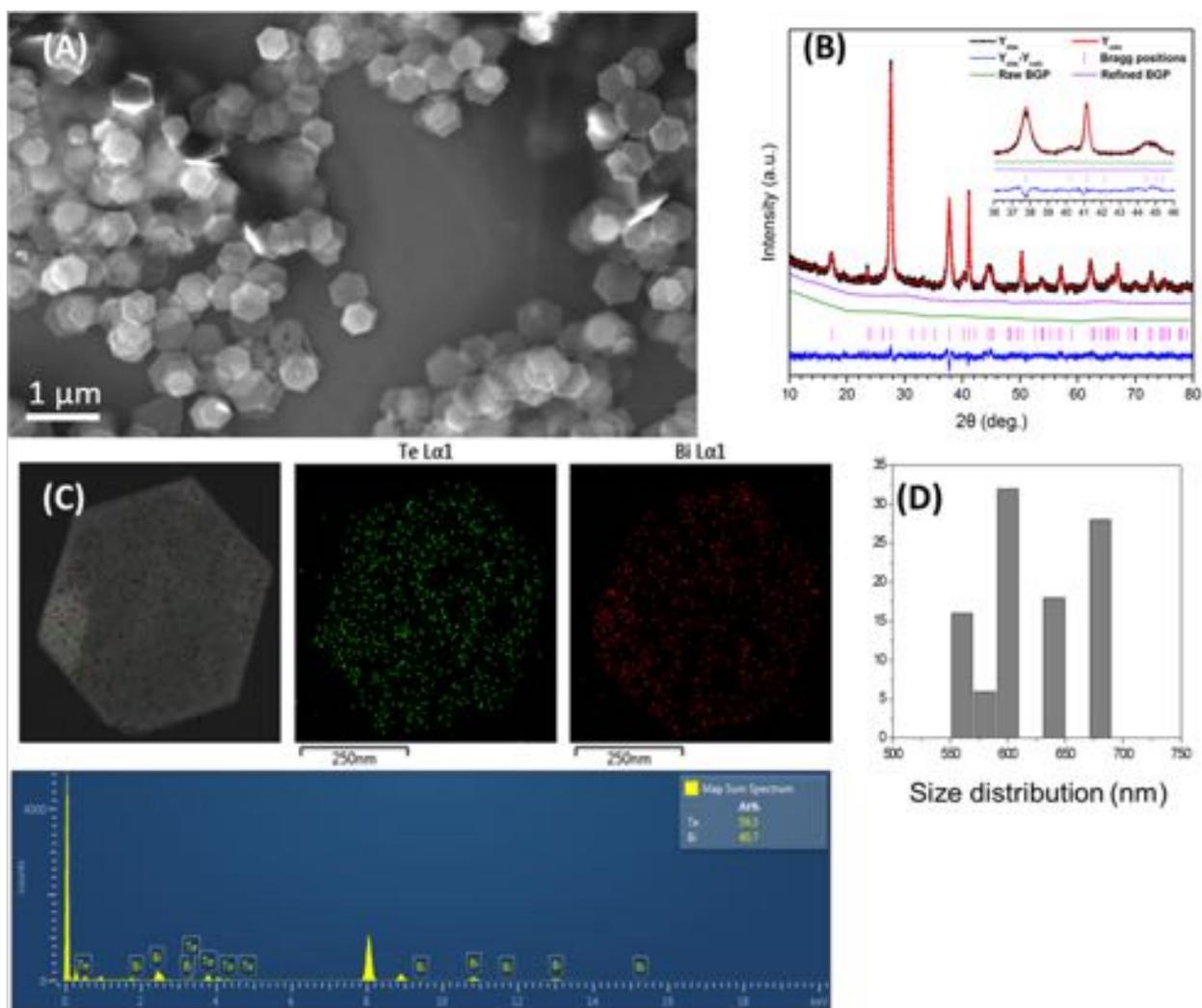

**Figure S1. Scanning Electron Microscopy (SEM) of Bi₂Te₃ nanoplatelets.** (A) Low-Magnification SEM image. (B) The experimental XRD pattern (black line) and the Rietveld analysis (red line) of the Bi₂Te₃ nanoplatelets. (C) Dark-field SEM with Te and Bi atomic mapping and EDX results with inset showing the atomic content percentage of Bi and Te. (D) Size distribution of the Bi₂Te₃ nanoplatelets.

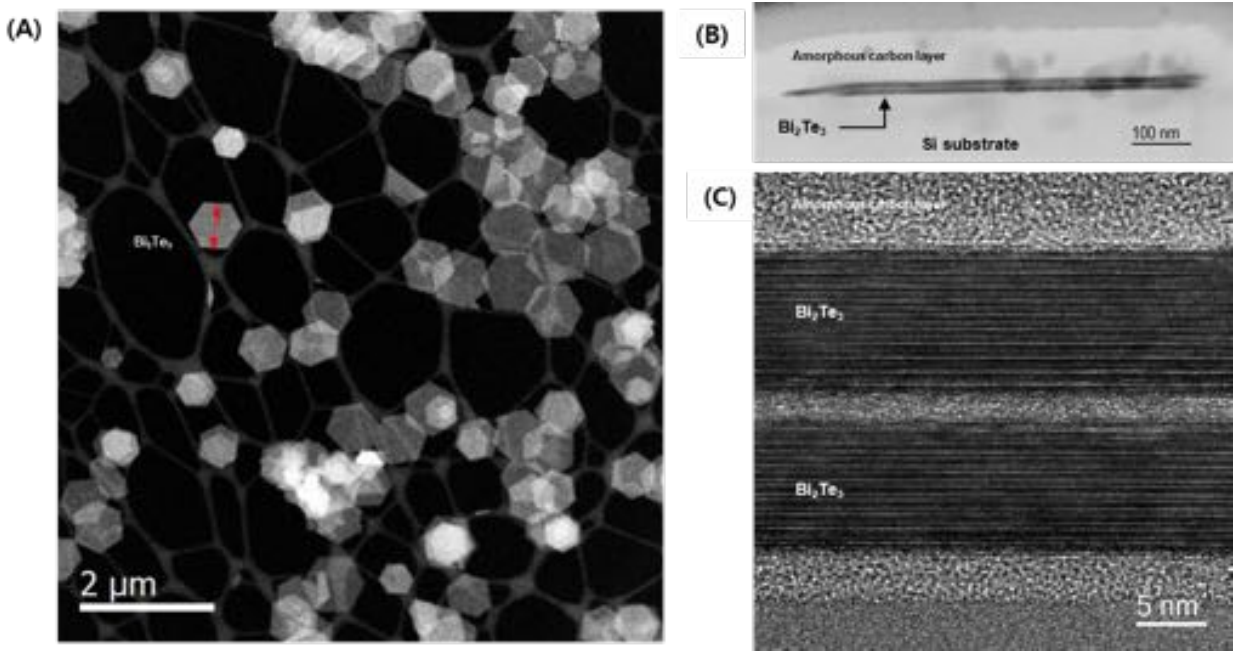

**Figure S2. Transmission Electron Microscope micrographs of Bi₂Te₃ nanoplatelets.**

(A) The low-magnification HAADF image shows the overall uniform size and hexagonal shape of the Bi₂Te₃ nanoplatelets. (B) A low-magnification TEM image in which the cross-section of two Bi₂Te₃ nanoplatelets is observed. (C) The magnified TEM image of the two Bi₂Te₃ nanoplatelets in Figure S2(B). The two Bi₂Te₃ nanoplatelets are separated by an amorphous layer (grey), and have thicknesses 10.4 nm and 9.5 nm, respectively.

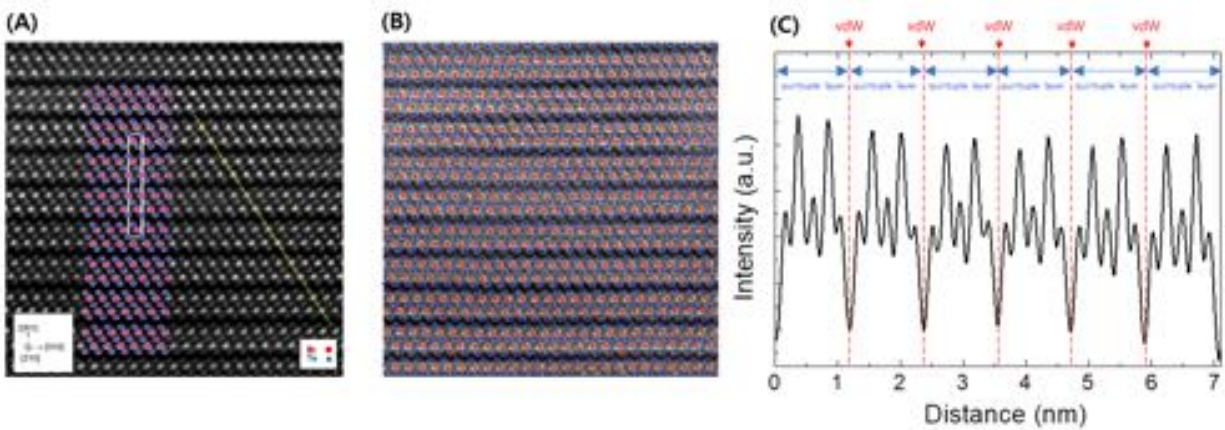

**Figure S3. Atomic position mapping of Bi₂Te₃ and intensity profiles of atomic columns.**

(A)   High-resolution cross-sectional HAADF image of Bi₂Te₃ along the [210] direction. The red and blue dots represent the Bi and Te columns, and the atomic model of the Bi₂Te₃ [210] direction overlays the HAADF image. (B) Chemical Analysis Image of the Bi₂Te₃ nanoplatelet. (C) Intensity profile along the yellow line in (A) shows the separation of the Bi and Te columns in quintuple layers. Van der Waals bonding is indicated between the quintuple layers.

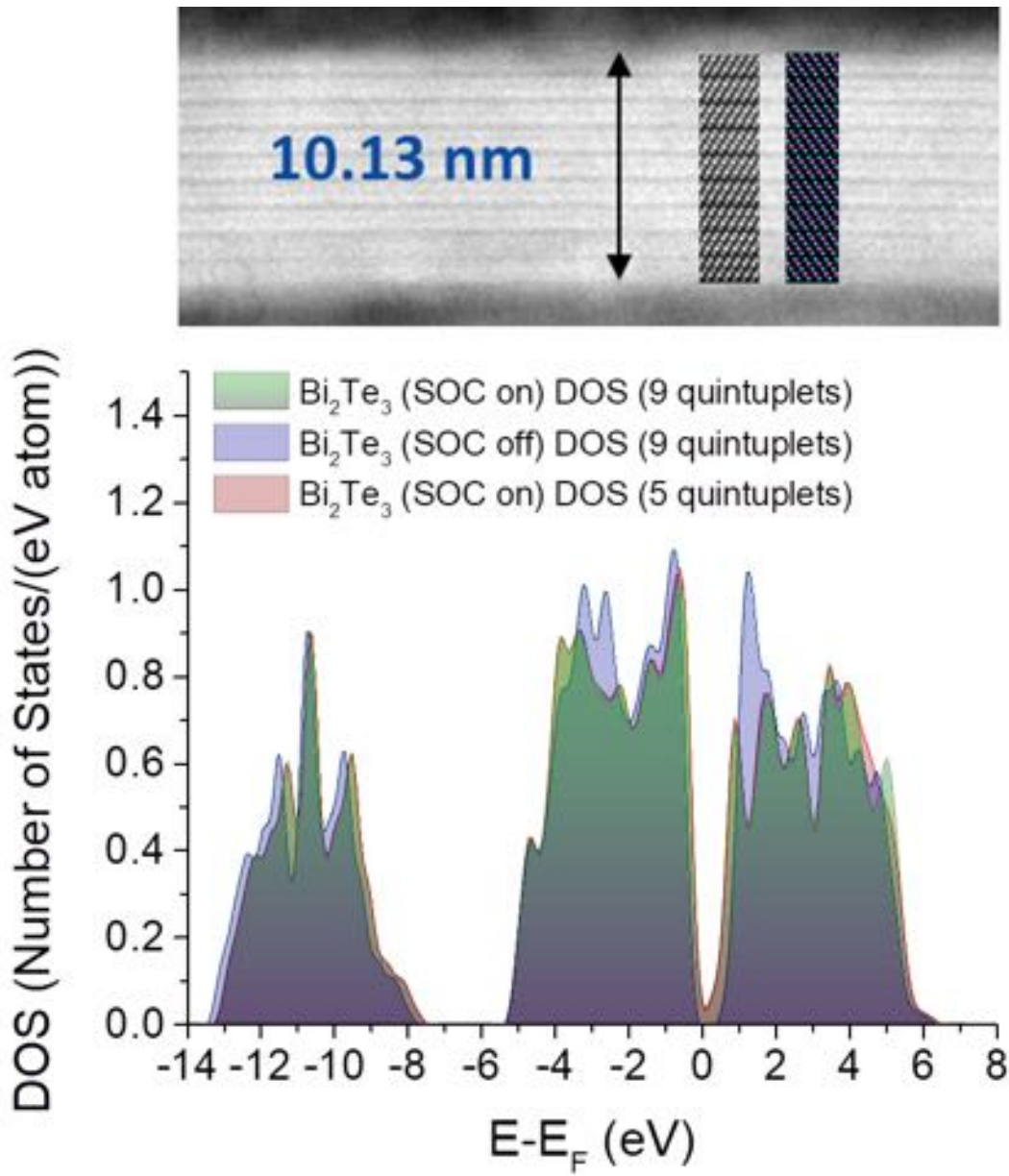

**Figure S4. DOS of Bi₂Te₃ slabs comprising both 5 quintuplets and 9 quintuplets calculated with and without SOC.** The 9-quintuplets slab corresponds to the average nanoplatelet thickness as seen in the upper panel of the Figure. When SOC is switched off an energy gap of ~200 meV opens at the Fermi level between the highest Valence Band and the Lowest Conduction Band. The extra DOS that is present in this gap when SOC is included is due to the surface Dirac states.

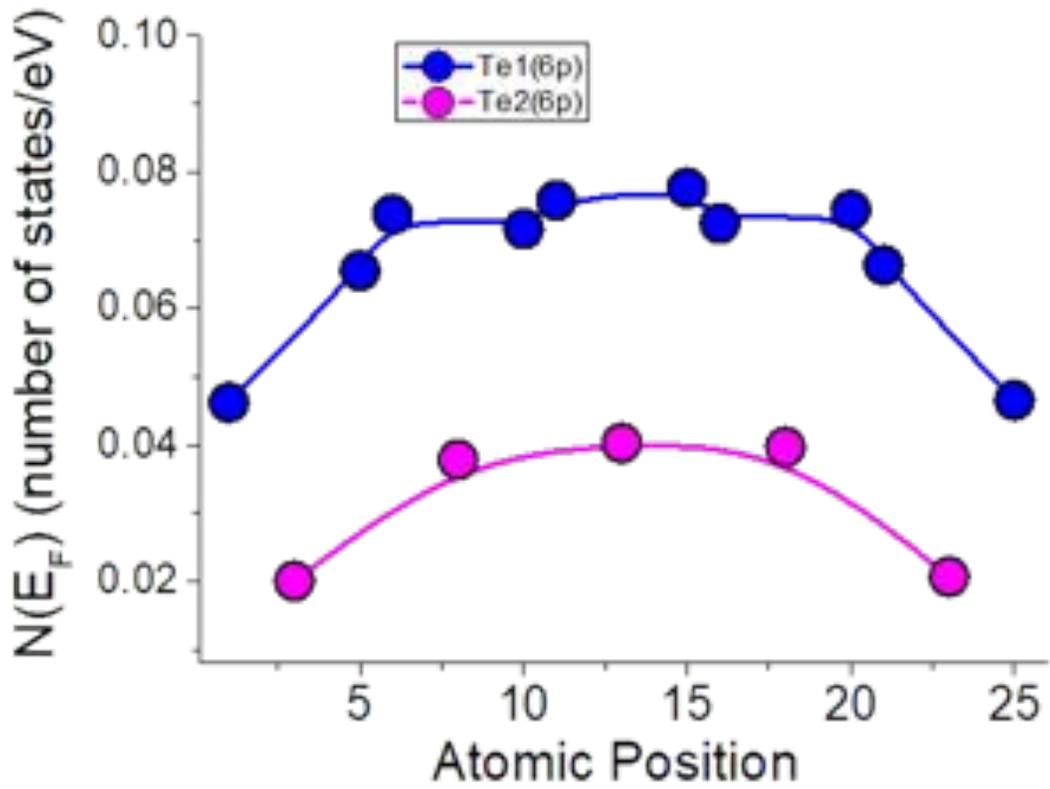

**Figure S5. The Te(1) and Te(2) DOS at the Fermi level $N(E_F)$, across a 5-quintuplet (~5 nm) Bi$_2$Te$_3$ slab.** Dirac electrons are observed to extend uniformly across the nanoplatelet.

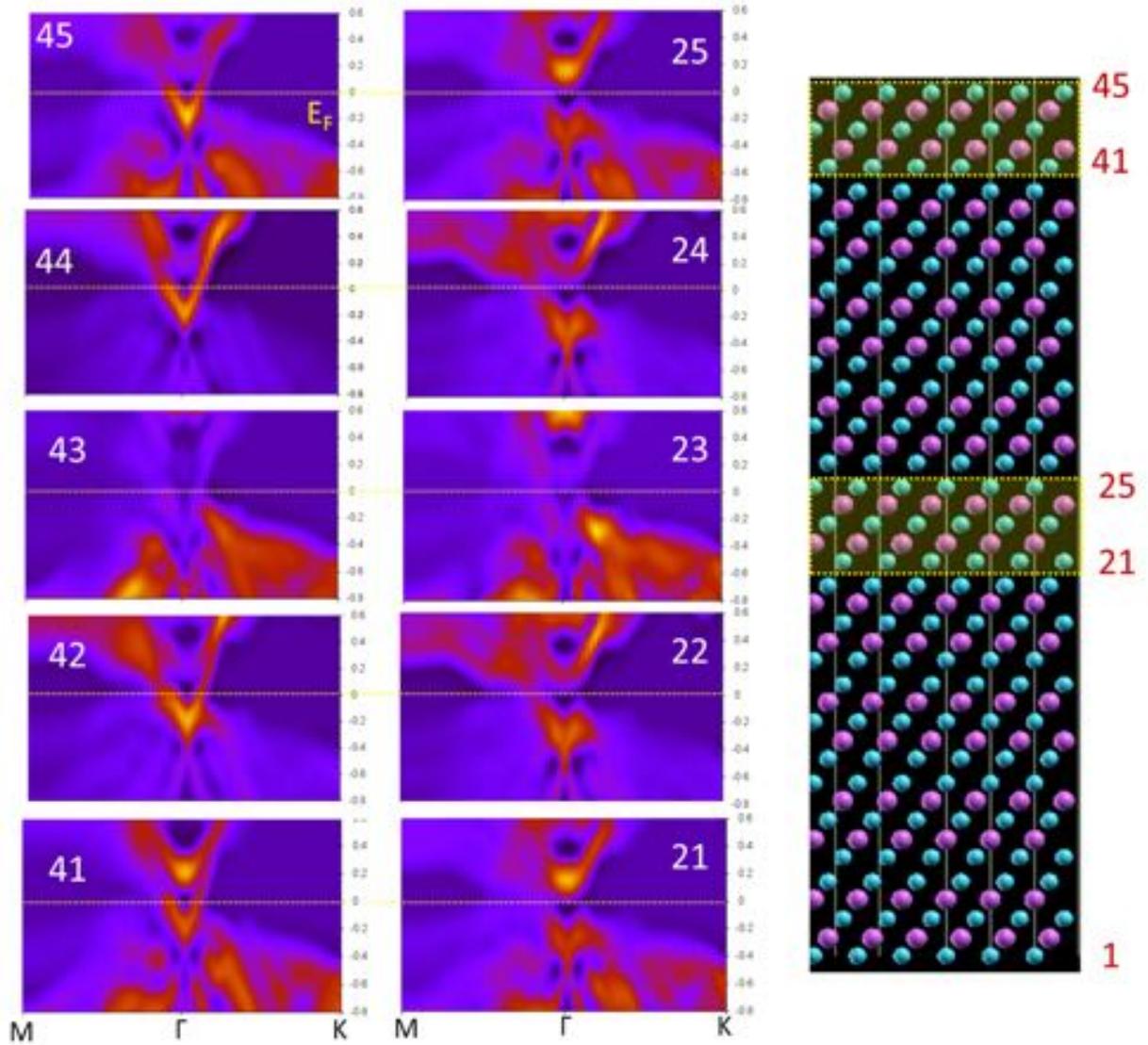

**Figure S6. *k*-resolved projected DOS in two different *k*-directions of a 9-quintuplets Bi₂Te₃ slab. The** cyan spheres in the right-hand image represent Te atoms, and the magenta spheres are Bi atoms. The atomic rows within the slab are numbered 1 to 45. The contribution of the central quintuplet (atomic rows 21-25) to the Dirac states is significantly reduced, as also shown in Figure 4B.

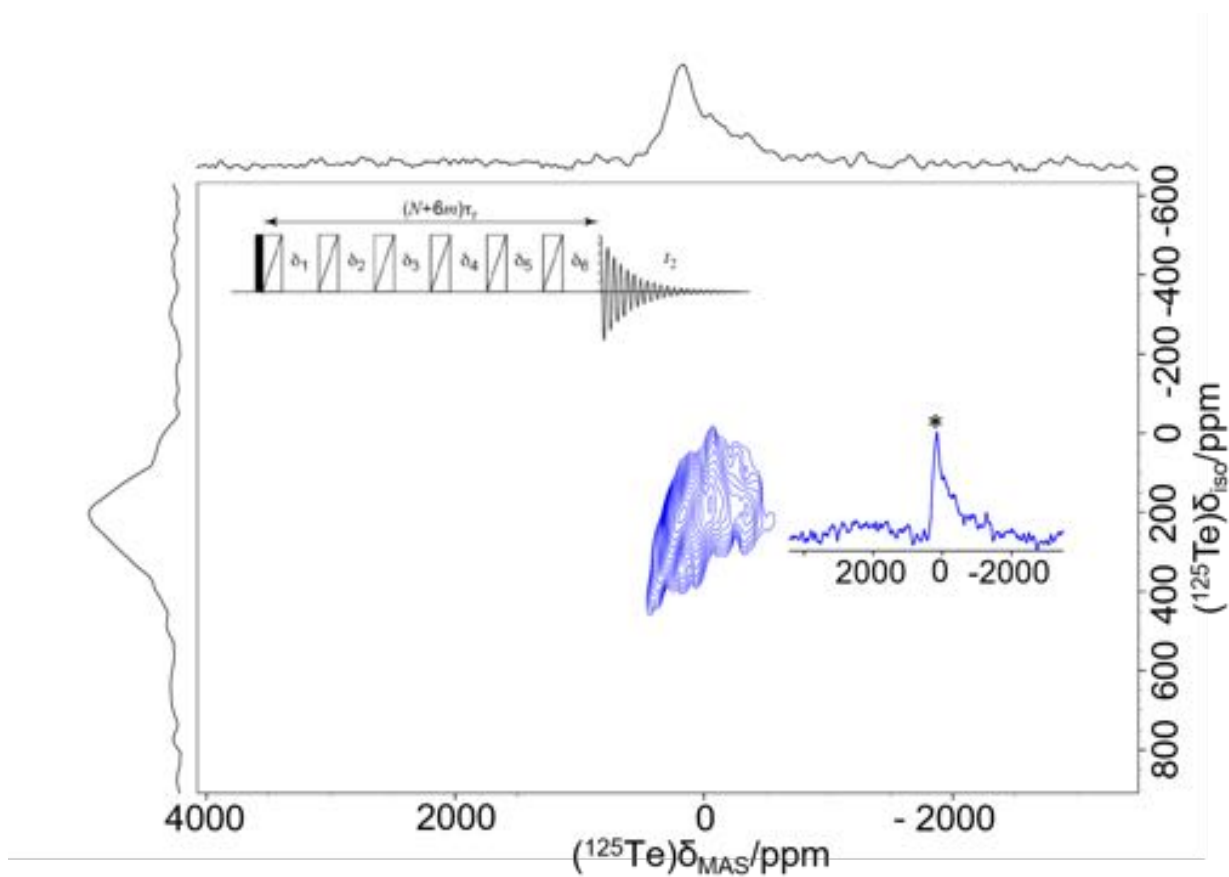

**Figure S7. $^{125}$Te aMAT NMR spectrum of bulk Bi$_2$Te$_3$.** Separation of the chemical shift and the chemical shift anisotropy is achieved in the bulk material. A single spectral feature, which comprises two overlapping resonances due to the two inequivalent Te sites, is observed at 250 ppm in the isotropic ($\delta_{iso}$) projection. Inset: The aMAT pulse sequence containing a 90° excitation pulse followed by six refocusing SHAPs prior to acquisition.

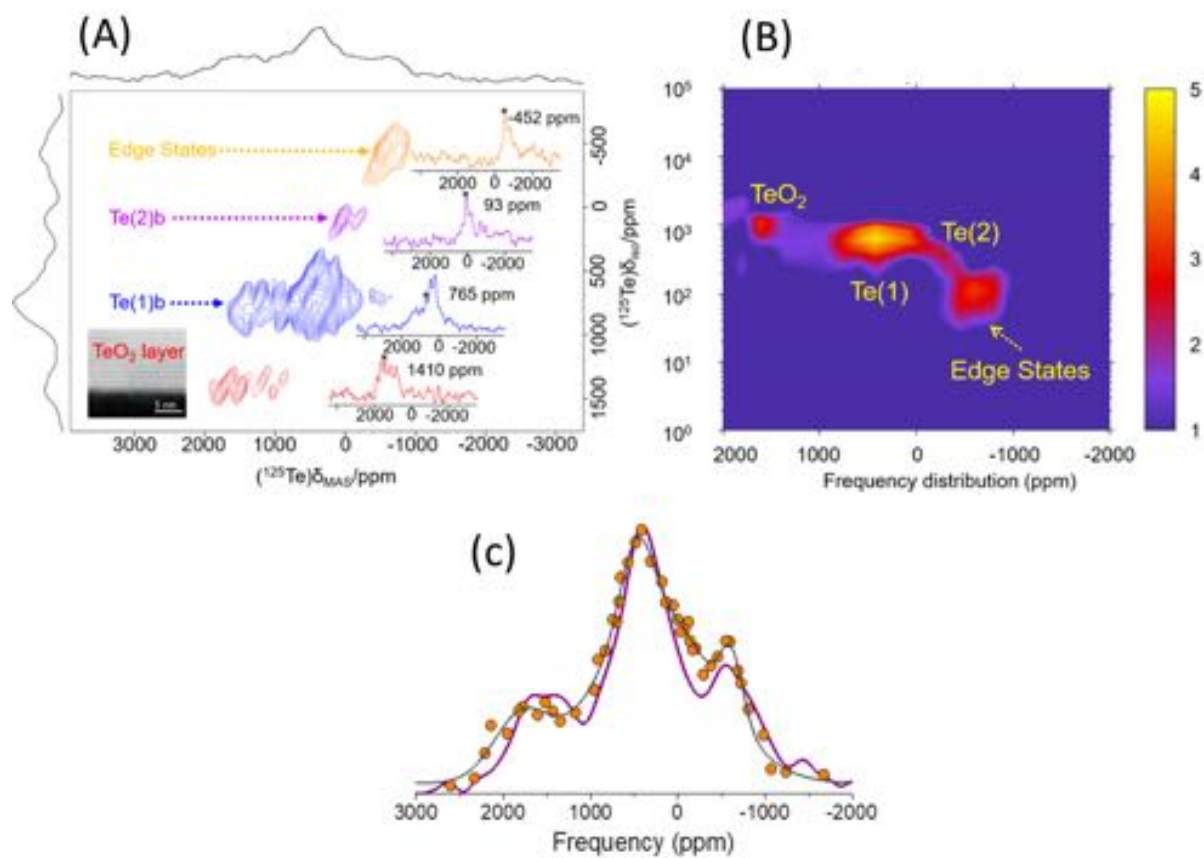

**Figure S8. ¹²⁵Te aMAT NMR, $T_2'$ distribution and Double Adiabatic Echo (DAE) of lightly-oxidized Bi₂Te₃ nanoplatelets.** (A) Fully relaxed aMAT spectrum, showcasing both the bulk and the edge states of the nanoplatelets, but also exposing a thin TeO₂ layer at 1410 ppm forming at the surface of a few nanoplatelets. (B) Fully-relaxed $T_2'$ distribution spectrum showcasing the same features. (C) The DAE (magenta color line) and the frequency sweep spectra (orange circles and black line).

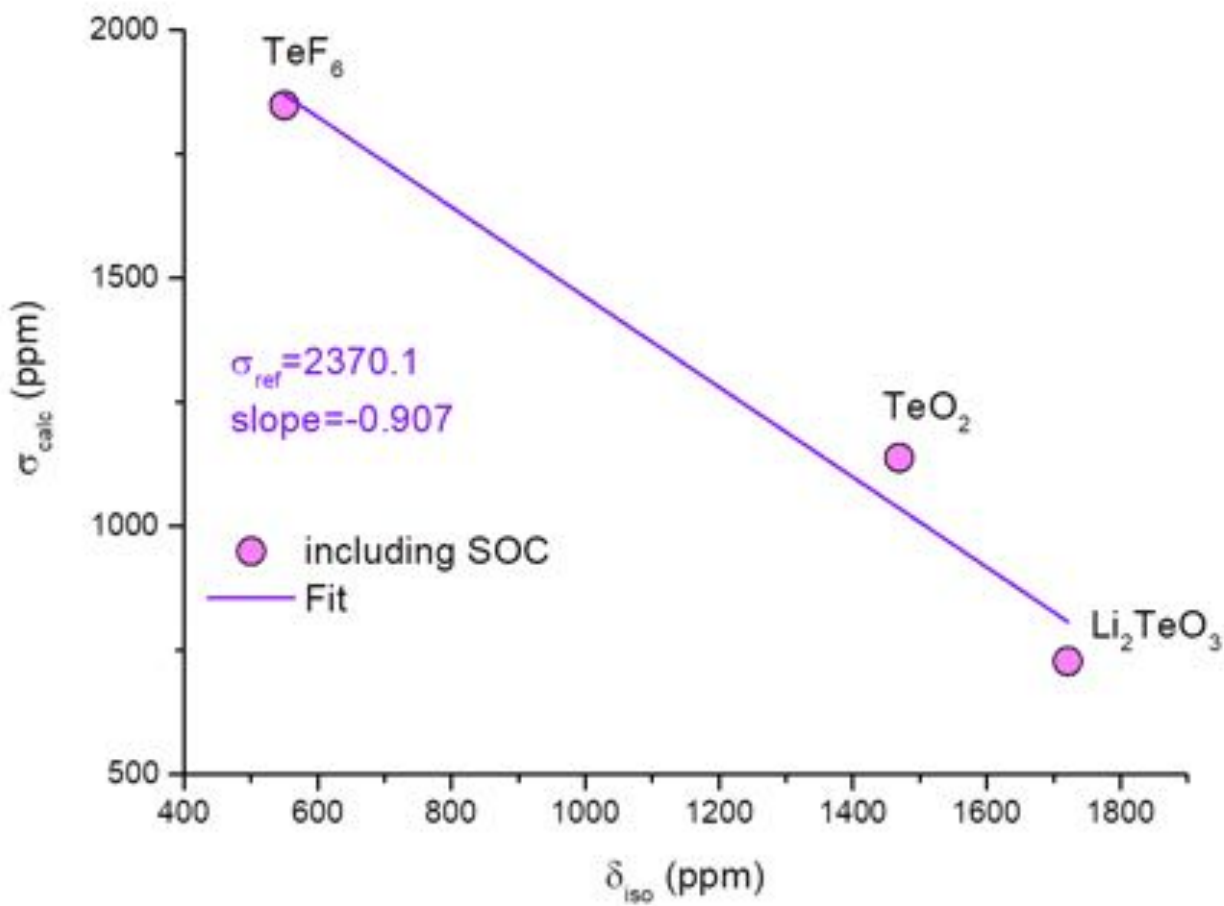

**Figure S9. Correlation between calculated $^{125}$Te NMR isotropic shielding and experimental isotropic chemical shift.** Data were considered for three representative compounds. According to the linear fit in the presence of SOC, the reference shielding is equal to $\sigma_{ref} = 2370.1$ ppm, and the slope is -0.907.

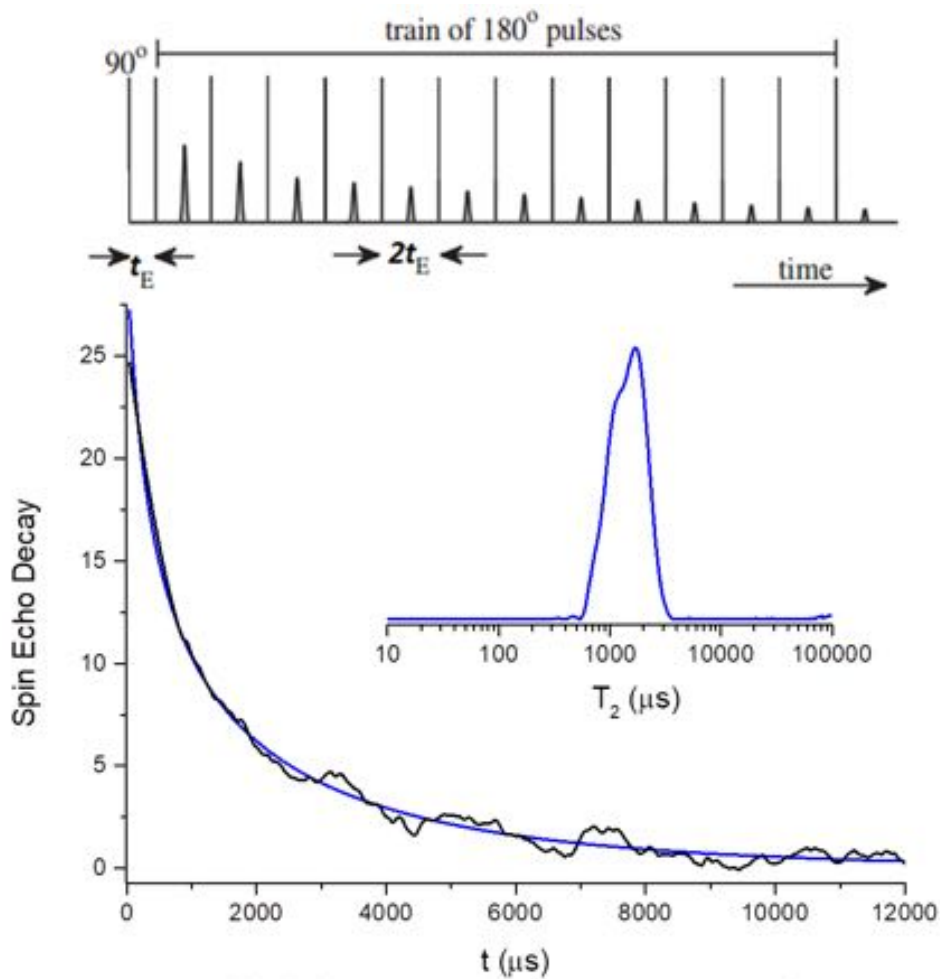

**Figure S10. The $^{125}$Te NMR $T_2'$ distribution (inset) obtained by inverting the experimental spin-echo decay (black curve in the main panel).** The experimental echo-train comprises 300 echoes with an interpulse separation of **$2t_E = 40$** μs. The inversion was performed by implementing a non-negative Tikhonov regularization algorithm. The blue curve in the main panel is the best monoexponential fit to the experimental spin-echo decay.

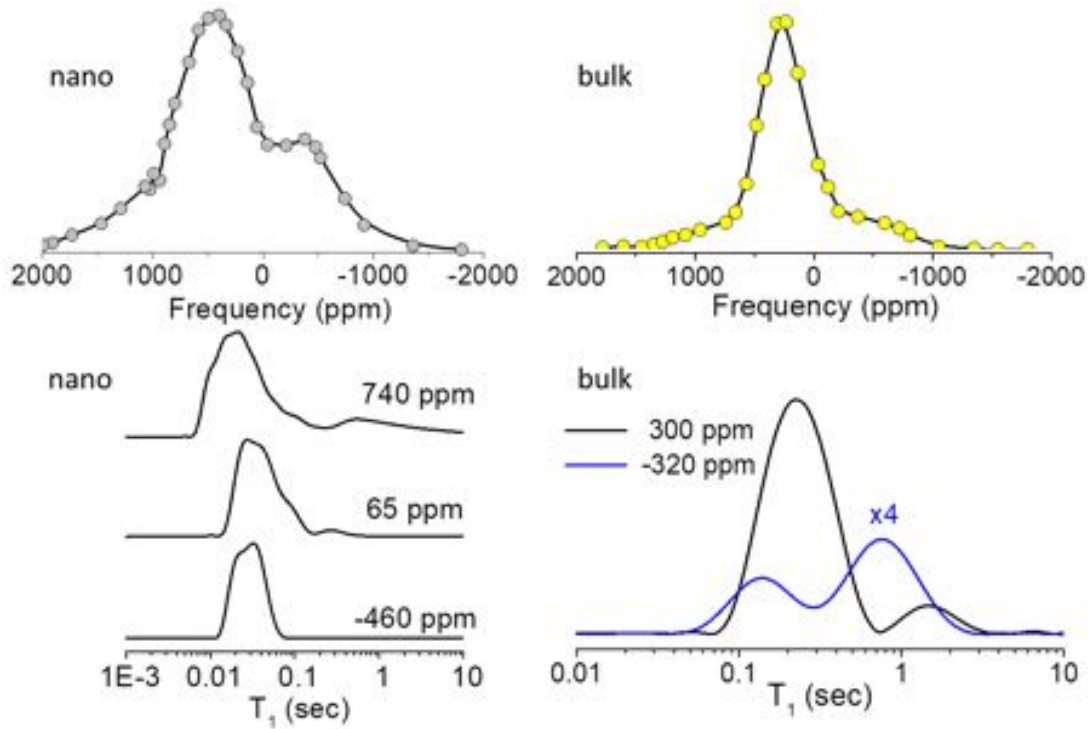

**Figure S11. The $^{125}$Te NMR spin-lattice relaxation time $T_1$ distribution of the nanoplatelets (left panels) and the bulk system (right panels) at characteristic frequencies.** For clarity, the signal intensities have been rescaled. In the case of the nanoplatelets the shown shifts are close to the three isotropic peaks observed in the aMAT spectrum. The $T_1$ values of the nanoplatelets are an order of magnitude shorter, reflecting the significant role of the Dirac electrons on the NMR relaxation processes across the entire nanoplatelet.